\documentclass[useAMS,usenatbib,usegraphicx]{mn2e}
\usepackage{amsmath,amssymb,amsbsy,amsfonts}
\usepackage{amssymb}
\usepackage[british]{babel}
\usepackage{mdwmath}
\usepackage{color}
\usepackage{url}

\title[Chemical history of isolated dwarf galaxies of the Local Group: I. dSphs: Cetus and Tucana]
{Chemical history of isolated dwarf galaxies of the Local Group: I. dSphs: Cetus and Tucana}
\author[N. Avila-Vergara, et al.]
 {N. Avila-Vergara,$^1$$^,$$^4$ \thanks{E-mail: navila@astro.unam.mx}
  L. Carigi$^1$, S. L. Hidalgo$^2$$^,$$^3$ and R. Durazo$^1$\\
  $^1$ Instituto de Astronom\'ia, Universidad Nacional Aut\'onoma de M\'exico, AP 70-264, 04510 M\'exico DF, M\'exico.\\
  $^2$ Instituto de Astrof\'isica de Canarias, V\'ia L\'actea s/n, E38200 La Laguna, Tenerife, Canary Islands, Spain.\\
  $^3$ Department of Astrophysics, University of La Laguna,  V\'ia L\'actea s/n, E38200 La Laguna, Tenerife, Canary Islands, Spain.\\
  $^4$ Departamento de F\'isica  y  Matem\'aticas, Universidad  Iberoamericana, Prolongaci\'on Paseo de la Reforma 880, Lomas de Santa Fe,\\ 
           CP 01210 M\'exico DF, M\'exico.\\}
\pagerange{\pageref{firstpage}--\pageref{lastpage}}

\begin{document}
%-----------------------------------------------------------------------------------------------------
\maketitle
\label{firstpage}
\begin{abstract}

For the first time, we obtain chemical evolution models (CEMs) for Tucana and Cetus, two isolated dwarf spheroidal galaxies (dSphs) of the Local Group. 
The CEMs have been built from the star formation histories (SFHs) and the metallicity histories, both obtained independently by the LCID project from deep color-magnitude diagrams. 
Based on our models, we find that the chemical histories were complex and can be divided into different epochs and scenarios. 
In particular, during 75\% of the SFH, the galaxies behaved as closed boxes and, during the remaining 25\%, either received a lot of primordial gas by accretion or they lost metals through metal-rich winds. 
In order to discriminate between these two scenarios, abundances ratios in old stars are needed. 
At t$\sim$4.5 Gyr, the galaxies lost most of their gas due to a short-strong, well-mixed wind.
We obtain very similar CEMs for both galaxies, although Cetus is twice as massive as Tucana.
We conclude that the star formation in both galaxies began with only 1.5\% of the baryonic mass fraction predicted by $\Lambda$CDM.

\end{abstract}

%--------------------------------------------------------------------------------
%--------------------------------------------------------------------------------
\section{Introduction}

The evolution of dwarf galaxies became a field of huge interest over the last decade, because dwarf galaxies are the most common galaxies in the local Universe and, 
in a hierarchical structure formation scenario, they are the first galaxies to form from which larger galaxies formed later on (Kravtsov et al. 1998).

Dwarf galaxies provide a small scale vision of the evolution of large galaxies in the early stages of the evolution.
In particular, the study of dwarf galaxies of the Local Group (LG) is fundamental, because the LG is the nearest cosmological laboratory to test theories on galactic evolution.

It is important to develop chemical evolution models (CEM) for dwarf galaxies, since it is through these models that we can infer the formation and evolution of the galaxies.
The CEMs are constructed based on several ingredients: star formation rate ($SFR$), initial mass function (IMF), stellar yields, primordial or enriched infalls and outflows. 
The models are constrained by the observed values of gaseous and stellar masses, chemical abundance ratios in HII regions, PNe and stars. 
Unfortunately, some of the ingredients of the CEMs and the observational constraints are not always known, diminishing the robustness of the model predictions.

However, for some dSphs of the LG, due to their proximity, stellar populations have been resolved and color-magnitude diagrams (CMDs) were obtained. 
From these CMDs the star formation histories were derived (e.g. Hernandez et al. 2000, Hidalgo et al. 2009, del Pino et al. 2013)
and some metallicity histories, or metal distribution functions were determined (e.g. Kirby et al. 2011a, Ross et al. 2015).

In the literature, we found several CEMS of dSphs of the LG. 
A few models took into account SFH obtained from CMDs (e.g. Carigi, Hernandez \& Gilmore 2002, Romano \& Starkenburg 2013), but the predicted metallicity histories could not be tested due to the lack of precise  metallicity histories. 
Some models inferred the SFH when reproducing the present-time properties and metal distribution functions (e.g. Lanfranchi \& Matteucci 2010, Romano \& Starkenburg 2013, Ross et al. 2015, for example). Other SFHs from CMDs and different CEMs for dSphs are mentioned in the excellent review article by Tolstoy, Hill, \& Tosi (2009).

Particularly, we are interested in isolated dSphs in order to study their evolution free of external factors, such as environmental interactions.
Therefore, our work is based on the star formation and metallicity histories from deep CMDs obtained by the LCID (Local Cosmology from Isolated Dwarfs) project (Monelli et al. 2010a, 2010b; Hidalgo et al. 2013; for example). 
This project analysed two dSphs: Tucana and Cetus, among six dwarf galaxies of the Local Group. 
A strong motivation for the study of this pair of galaxies is to discriminate between nurture versus nature frames: if these galaxies have a common origin or they were different from their origin.

In this work, and for the first time in the chemical evolution field, we focus on developing CEMs for Tucana and Cetus.  
In our opinion, these dwarf galaxies are truly isolated, because they are currently located outside the virial radii of Andromeda and the Milky Way galaxies (McConnachie 2012).
It is important to note that this is a novel work, because our chemical evolution models consider, for each evolution time, the star formation histories (as a chemical ingredient) and the metallicity history (as an observational constraint).

%---------------------------------------------------------------------------------------------------------------------
\section{Observational constraints}

Our chemical evolution models for two isolated dSph galaxies of the Local Group, Tucana and Cetus, have been built based on star formation histories from CDM (see Figure 1) to reproduce the following observational constraints: the current mass of gas and mass of all stars ever formed (see Table 1) and the metallicity history (see Figure 2). 

There are few observational data of gaseous mass for both galaxies. 
In the literature, we found only one value for Tucana, which is $M_{gas}^{Obs} = 1.5 \times10^{4}M_\odot$ by Oosterloo (1996) and none for Cetus.

The star formation rate vs time, $SFR$(t), hereafter star formation history (SFH), is one of the most important ingredients for constructing a CEM. 
We compute the global $SFR_{field}$(t), from the radial variation of the star formation history, $SFR_{field}$(r,t), by Hidalgo et al. (2013, H13).
In particular, $SFR_{field}(t) = \int SFR_{field}(r,t) dA$, where $dA$ is the differential area. 
The integration was calculated for the four elliptical regions observed by H13. 
From this $SFH_{field}$, we compute the mass of all stars ever formed $M_{*field}^{formed} = \int _{0}^{t} SFR_{field}(t)dt$ and 
obtain $M_{*formed}^{field}(Tucana)= 2.77\times10^6M_\odot$,  $M_{*formed}^{field}(Cetus)=1.82\times10^6M_\odot$. 
We also calculate the mass of living stars, $M_*^{field}= (1-R)\int_{0}^{t} SFR_{field}(t)dt$ (see eq. 2) and obtain $M_*^{field}(Tucana)= 2.37 \times10^6M_\odot$, $M_*^{field}(Cetus)= 5.19 \times10^6M_{\odot}$. 
In both galaxies the global $SFH_{field}$ and the $M_*^{field}$ are in agreement with the global $SFR$(t) and stellar mass by Monelli et al. (2010 a,b). 

Hidalgo et al. (2013) also obtained the mass of all stars ever formed for the whole galaxies ($M_*^{formed}$, see their table 2) from the $SFR_{field}$(r,t).
They extended the $SFR_{field}$(r,t) to r infinite and assumed: i) that the observed field (r $<$ $R_{4}$) is a good sample of the rest of each galaxy (r $>$ $R_{4}$), and ii) the radial profile ($\propto e^{-r/\alpha_{\psi}}$) of stellar mass density follows the same radial decline beyond the maximum observed radius ($R_{4}$). 
Particularly, $R_{4}^{Tucana}=5.9 \alpha_{\psi}$ and $R_{4}^{Cetus}=4.5 \alpha_{\psi}$, where $\alpha_{\psi}$=121pc and 212 pc for Tucana and Cetus, respectively. 
It is important to remark that the all figures about \textit{SFR} shown in H13 correspond to the observed field, that is $SFR_{field}$.

In order to obtain the \textit{SFR}(t) for the whole galaxy, which are considered in the CEMs, we increase the global $SFR_{field}(t)$ by a scaling factor. 
Specifically, this factor is equal to $M_{*}^{formed}/ M_{*formed}^{field}$ and its value is 1.15 for Tucana and 3.84 for Cetus. 
In Figure 1 we show the SFHs for the entire galaxies.

We also computed the global metallicity histories, \textit{Z}(t), $Z(t) = \frac{ \int Z(r,t)SFR(r,t)dA}{SFR(t)}$, where \textit{Z}(r,t) is the radial variation of the metallicity history by H13. 
The integration was calculated for the four elliptical regions observed by H13. 
In both galaxies, $Z(t)$ is in good agreement with Monelli et al. (2010 a,b). 
In Figure 2 the metallicity histories for both galaxies are present. 
Each stellar metallicity value ($Z_{*}$) corresponds to the 1-Gyr-average metallicity of stars formed during this Gyr. 
Moreover each $Z_{*}$ value represents the 1-Gyr-average metallicity of the gas, because $Z_{*}$ corresponds to the metallicity of the stars formed with the gas present at that time.
It is relevant to note that these $Z_{*}$ values were obtained from synthetic CMDs.
In particular for Tucana, the two first $Z_{*}$ values $\sim$($4.89\pm1.22$)$\times10^{-4}$, between 0 and 2 Gyr, agree with the mean [Fe/H] value determined from Ca triplet absorption lines in red giant branch stars by Fraternali et al. (2009). 
They found [Fe/H] = -1.95$\pm$0.15, which corresponds to $Z_{spec}=$($2.38\pm0.79$)$\times10^{-4}$.
Similarly for Cetus, the $Z_{*}$(0-2Gyr)=($3.30\pm1.00$) $\times10^{-4}$ is in agreement to the Fe abundance obtained by Lewis et al. (2007). 
They estimated [Fe/H] $\sim$ -1.90, which is equivalent to $Z_{spec} \sim 2.52\times10^{-4}$. In both cases, the spectroscopic metallicities are similar to the $Z_{*}$ for t$<$2 Gyr, when the \textit{SFR} is higher, but $Z_{spec}$ is lower than $Z_{*}$ for t $>$2 Gyr, when the \textit{SFR} is lower.

Commonly, abundance ratios of different heavy elements ([$X_{i}/X{j}$], i$\neq$j$\neq$H) are used as observational constraints in CEMs for dSph galaxies (Carigi et al. 2002; Lanfranchi \& Matteucci 2010; Romano \& Starkenburg 2013, for example). 
Unfortunately, we cannot use this type of observational data, because in the literature there is no available data of [$X_{i}/X{j}$]  for Tucana and Cetus.

%"""""""""""""""""""""""""""""""""""""""""""""""""""""""""""""""""""""""""""""""""""""""""""""""""""""""
\begin{table}
 \caption{Observational constraints.}
 \label{table1}
 \begin{tabular}{@{}lcccccc}
  \hline
   & Tucana   &  Cetus\\
  \hline
 $M_{gas}^{Obs}(10^4M_\odot)$ & $1.5^a$ & $-$\\ 
 $M_*^{formed}(10^6M_\odot)$ & $3.2^b$  & $7.0^b$\\
 $M_{vir}(10^9M_\odot)$ & $4^c$ & $-$\\
 $Z_*(10^{-3})$ & $1.38\pm0.23^d$ & $1.08\pm0.19^d$\\
 \hline
 \end{tabular}
 \medskip\\
 %Column 1: Observational constraints. Column 2 and 3: Name of galaxy.
 ${a}$ Current mass of gas, Oosterloo et al. (1996).\\
 ${b}$ Mass of all stars ever formed, Hidalgo et al. (2013, H13).\\
 ${c}$ Virial mass,  Monelli et al. (2010a).\\
 ${d}$ Average stellar metallicity at 5.5 Gyr, H13, see Section 2.\\
 \end{table}
%"""""""""""""""""""""""""""""""""""""""""""""""""""""""""""""""""""""""""""""""""""""""""""""""""""""""
%-------------------------------------------------------------------------------------------------------------------------------------------
\section{Equations and general assumptions of the CEMs}

The equations that describe the chemical evolution of the galaxies are integrodifferentials, when considering the lifetime of each star (e. g. Tinsley 1980). 
In order to solve these equations analytically, we adopt the Instantaneous Recycling Approximation (IRA). 
This approximation assumes that the lifetimes of the stars more massive than 1$M_\odot$ are negligible when compared to the galactic age.
 In some chemical evolution models the adoption of IRA for $m>1 M_\odot$ is common, in particular when CEMs are built to reproduce observational constraints regarding to O. 
 We consider that the IRA is a very good approximation for our models, because: 
 i) the metallicity histories, that models try to reproduce, were obtained from the initial metallicities used in the synthetic CDMs, 
 ii) O is the most abundant heavy element included in \textit{Z}, and 
 iii) there are no abundance determinations of alpha elements for Tucana and Cetus.

%NO SE COMO PONER UN ESPACIO AQUI
 The evolutionary equation of baryonic mass ($M_{bar}$) is
  %"""""""""""""""""""""""""""""""""""""""""""""""""""""""""""""""""""""""""""""""""""""""""""""""""""""""
\begin{equation}
	\frac{dM_{bar}(t)}{dt}=\frac{d(M_{*}(t)+M_{gas}(t))}{dt}=A(t)-W(t),
\end{equation}
%"""""""""""""""""""""""""""""""""""""""""""""""""""""""""""""""""""""""""""""""""""""""""""""""""""""""
where $M_{gas}$ and $M_*$ represent gas mass and stellar mass, respectively.

When IRA is assumed
\begin{equation}
	\frac{dM_{*}(t)}{dt}=(1-R)SFR(t),
\end{equation}
%"""""""""""""""""""""""""""""""""""""""""""""""""""""""""""""""""""""""""""""""""""""""""""""""""""""""
\begin{equation}
	\frac{dM_{gas}(t)}{dt}=-(1-R)SFR(t)+A(t)-W(t), 
\end{equation}
%"""""""""""""""""""""""""""""""""""""""""""""""""""""""""""""""""""""""""""""""""""""""""""""""""""""""
where \textit{A}, \textit{W}, \textit{SFR} and \textit{R} represent the infall, outflow, star formation rates and the fraction of the mass ejected to the interstellar medium (ISM) by the death of stars.
Moreover, the evolution of metallicity, $Z(t)$, is as follows:

%"""""""""""""""""""""""""""""""""""""""""""""""""""""""""""""""""""""""""""""""""""""""""""""""""""""""
\begin{equation} 
\begin{aligned}
	\frac{d(Z(t)M_{gas}(t))}{dt}=&-Z(t)(1-R)SFR(t)+(1-R)Y_{Z}SFR(t)\\
	&+Z_{A}(t)A(t)-Z_{W}(t)W(t),
\end{aligned}
\end{equation}
%"""""""""""""""""""""""""""""""""""""""""""""""""""""""""""""""""""""""""""""""""""""""""""""""""""""""
where
$Y_{Z}$ is a fraction of the metals synthesized and ejected by the death of stars and $Z_{A}$ and $Z_{W}$ are the infall and outflow metallicities.
$R$ and $Y_{Z}$ can be expressed as the sum of the contributions of low-intermediate mass stars (LIMS,  $1-7.5M_{\odot}$) and massive stars (MS, $7.5M_{\odot}-M_{up}$) as follows:
%"""""""""""""""""""""""""""""""""""""""""""""""""""""""""""""""""""""""""""""""""""""""""""""""""""""""
\begin{equation}
	 R=R_{LIMS}+R_{MS},
\end{equation}
%"""""""""""""""""""""""""""""""""""""""""""""""""""""""""""""""""""""""""""""""""""""""""""""""""""""""
\begin{equation}
	Y_{Z}=\frac{P_{Z,LIMS}+P_{Z,MS}}{1-R},
\end{equation}
%"""""""""""""""""""""""""""""""""""""""""""""""""""""""""""""""""""""""""""""""""""""""""""""""""""""""
where $R_{LIMS}$, $R_{MS}$ are the returned masses, and $P_{Z,LIMS}$, $P_{Z,MS}$ are the integrated yields, for LIMS and MS respectively.

%"""""""""""""""""""""""""""""""""""""""""""""""""""""""""""""""""""""""""""""""""""""""""""""""""""""""
%\begin{equation}
%	P_{Z,LIMS}=\int _{1 M_{\odot}}^{7.5 M_{\odot}} mp_{z}(m)\phi(m)dm
%\end{equation}
%"""""""""""""""""""""""""""""""""""""""""""""""""""""""""""""""""""""""""""""""""""""""""""""""""""""""
%\begin{equation}
%	P_{Z,MS}=\int _{7.5 M_{\odot}}^{M_{up}} mp_{z}(m)\phi(m)dm
%\end{equation}
%"""""""""""""""""""""""""""""""""""""""""""""""""""""""""""""""""""""""""""""""""""""""""""""""""""""""
Below, we describe the general assumptions for the CEMs shown in this paper.\\
i) Instantaneous recycling approximation.\\
ii) The virial mass, $M_{vir}$, is constant over time.\\
iii) The initial baryonic mass, $M_{bar}(0)$, is a fraction $f_{b}(0)$ of $M_{vir}$, $f_{b}(0)=\frac{M_{bar}(0)}{M_{vir}}$.\\
iv) The initial baryonic mass is only formed by gas, $M_{bar}(0)\equiv M_{gas}(0)$.\\
iv) $R_{MS}$, $R_{LIMS}$, $P_{Z,MS}$ and $P_{Z,LIMS}$ are taken from Hern\'andez-Mart\'inez et al. (2011), they considered IMF by Kroupa et al. (1993). 
We choose the values for $Z_{i} = 4.0\times  10^{-3}$ and $M_{up}=40M_{\odot}$, obtaining $R=0.258$ and $Y_{Z}=9.456 \times 10^{-3}$.\\
v) The star-formation rate is a time function given by Hidalgo et al. (2013), see section 2 and Fig. 1.\\
%"""""""""""""""""""""""""""""""""""""""""""""""""""""""""""""""""""""""""""""""""""""""""""""""""""""""
vi) The galactic time, $t_{g}$, is equal to 13.5 Gyr.\\

Based on the previous assumptions, first we develop simple chemical evolution models, considering in each model only one galactic stage throughout the whole galactic lifetime. 

The simple CEMs are:

A) The closed-box model (CBM). This model considers that no gas escapes from or is added to the galaxy, i.e. $A(t)=W(t)=0$. Then the evolution of $M_{bar}$, $M_{gas}$ and \textit{Z} are written as:
%"""""""""""""""""""""""""""""""""""""""""""""""""""""""""""""""""""""""""""""""""""""""""""""""""""""""
\begin{eqnarray}
		M_{bar}(t)=M_{bar}(0)=M_{gas}(0),\\
		M_{gas}(t)=M_{gas}(0)-(1-R)\int_{0}^{t}SFR(t)dt,\\
		Z(t)=-Y_{Z}ln[\mu(t)],
\end{eqnarray}
%"""""""""""""""""""""""""""""""""""""""""""""""""""""""""""""""""""""""""""""""""""""""""""""""""""""""
where
%"""""""""""""""""""""""""""""""""""""""""""""""""""""""""""""""""""""""""""""""""""""""""""""""""""""""
\begin{equation}
	\mu(t)=\frac{M_{gas}(t)}{M_{bar}(0)}.
\end{equation}
%"""""""""""""""""""""""""""""""""""""""""""""""""""""""""""""""""""""""""""""""""""""""""""""""""""""""

B) The primordial accretion model (PAM) assumes that the galaxy is formed by accretion, $A(t)\neq 0$, of primordial material, $Z_{A}(t)=0$, but no gas escapes from the galaxy, $W(t)= 0$. 
The accretion rate is proportional to the \textit{SFR}, i.e, $A(t)=a(1-R)SFR(t)$, then the evolution of $M_{bar}$, $M_{gas}$ and \textit{Z} are as follow:
%"""""""""""""""""""""""""""""""""""""""""""""""""""""""""""""""""""""""""""""""""""""""""""""""""""""""
\begin{eqnarray}
		M_{bar}(t)=M_{bar}(0)+a(1-R)\int_{0}^{t}SFR(t)dt,\\
		M_{gas}(t)=M_{gas}(0)-(1-a)(1-R)\int_{0}^{t}SFR(t)dt,\\
		Z(t)=\frac{Y_{Z}}{a}[1-\mu(t)^{\frac{a}{1-a}}],
\end{eqnarray}
%"""""""""""""""""""""""""""""""""""""""""""""""""""""""""""""""""""""""""""""""""""""""""""""""""""""""
where, $a$ is the accretion efficiency and $\mu(t)$ is as in eq. 10.\\

C) The well mixed galactic wind model (WWM) considers that the galaxy loses gas $W(t)\neq 0$, but gas does not fall into the galaxy, $A(t)=0$. 
We assume that the material ejected by MS and the LIMS is well mixed with the interstellar medium before the ISM is expelled to the intergalactic medium, consequently $Z_{W}(t)=Z(t)$. 
The outflow rate is proportional to the \textit{SFR}, i.e, $W(t)=w(1-R)SFR(t)$, then the evolution of $M_{bar}$, $M_{gas}$ and \textit{Z} is given by:
%"""""""""""""""""""""""""""""""""""""""""""""""""""""""""""""""""""""""""""""""""""""""""""""""""""""""
	\begin{eqnarray}
		M_{bar}(t)=M_{bar}(0)-w(1-R)\int_{0}^{t}SFR(t)dt,\\
		M_{gas}(t)= M_{gas}(0)-(1+w)(1-R)\int_{0}^{t}SFR(t)dt,\\
		Z(t)=-\frac{Y_{Z}}{(1+w)}ln[\mu(t)],
	\end{eqnarray}
%"""""""""""""""""""""""""""""""""""""""""""""""""""""""""""""""""""""""""""""""""""""""""""""""""""""""
where, $w$ is the wind efficiency and $\mu(t)$ is as eq. 10.\\

D) The reduced yield model (RYM) adopts the same equations of the CBM, but with a reduced yield. 
In this model the original yield is divided by a factor e,  $e>1$, in the equations of the CBM. Therefore

\begin{eqnarray}
		M_{bar}(t) = cte,\\
		M_{gas}(t)= M_{gas}(0)-(1-R)\int_{0}^{t}SFR(t)dt,\\
		Z(t) = -\frac{Y_{Z}}{e}ln[\mu(t)].
\end{eqnarray}\\

A reduced yield may be due to the following two main processes:   

    1) Metal rich winds. When the \textit{SFR} is high, the number of supernovae (SNe) is high and the OB associations in their last stages of evolution can create a galactic bubble. Through this bubble a percentage of the material produced by massive stars, and even by the intermediate mass star, is expelled to the Intergalactic medium without previously mixing with the ISM (Rodr\'iguez-Gonz\'alez et al. 2011). Therefore, this ejected material does not contribute to the chemical enrichment of the galactic ISM.\\
    
    2) Stochastic effects. When the \textit{SFR} is small, the number of very massive stars formed can be low. 
    This produces stochastic effects in the IMF, specifically in the massive stars range. 
    Therefore the ISM enrichment is due mainly to the lower \textbf{massive} stars. 
    This stochasticity produces dispersions in the abundance ratios and a reduction in the O value (see figs 10 and 11 by Carigi \& Hernandez 2008).\\\\

The \textit{Z(t}) equations shown in this section are expressed in terms of $\mu(t)=\frac{M_{gas}(t)}{M_{bar}(0)}$, while the canonical \textit{Z(t)} equations are expressed in terms of $\mu(t)=\frac{M_{gas}(t)}{M_{bar}(t)}$, because  $M_{bar}(0)$ is a free parameter of our models and our equations are more compact than the canonical ones (e.g. Tinsley 1980, Matteucci 2001).

\begin{figure}
\begin{center}
\includegraphics[scale=0.48]{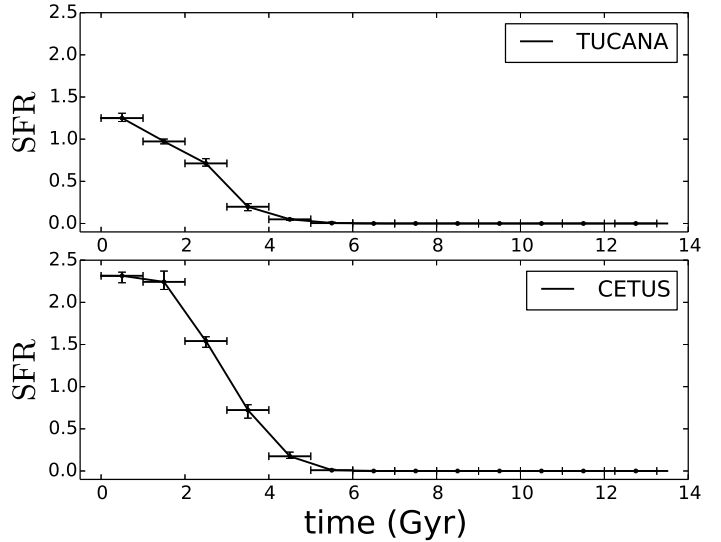}
\end{center}
\caption[Alphavsx]
{
Evolution of star formation rate (\textit{SFR}, $10^{6} M_{\odot} Gyr^{-1}$) for Tucana and Cetus. 
Points and uncertainties by H13 (see section 2) from CMD. 
$SFR$(t) is equivalent to the star formation history (SFH).}

\label{fig01}
\end{figure} 
%"""""""""""""""""""""""""""""""""""""""""""""""""""""""""""""""""""""""""""""""""""""""""""""""""""""""
\begin{figure}
\begin{center}
\includegraphics[scale=0.48]{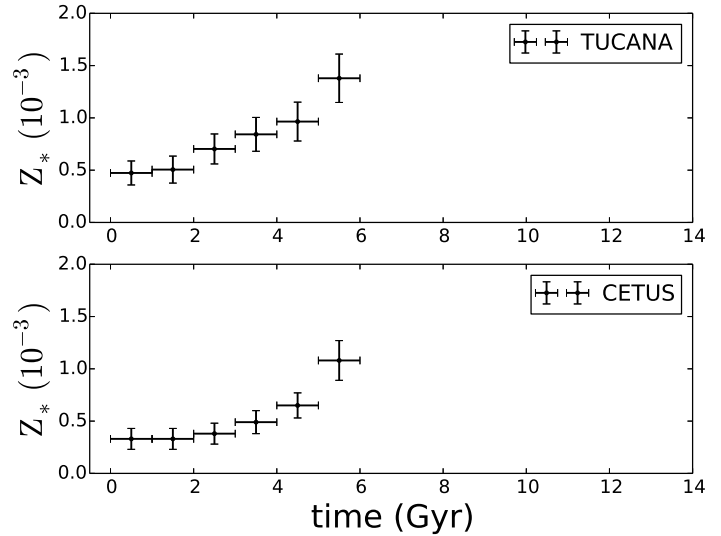}
\end{center}
\caption[Alphavsx]{
Evolution of the average stellar metallicity ( $Z_*$) for Tucana and Cetus. 
Points and uncertainties by H13 (see section 2) from CMD.
$Z_*(t)$ is equivalent to the metallicity history ($Z(t)$).
}
\label{fig02}
\end{figure}
%"""""""""""""""""""""""""""""""""""""""""""""""""""""""""""""""""""""""""""""""""""""""""""""""""""""""
\section{CEMs for Tucana}

\subsection{Simple CEMs for Tucana}

{Analysing the time behaviour of \textit{SFR} and \textit{Z} obtained from CMD (see Figs 1 and 2),
we think that simple CEMs, that assume only one scenario for the entire lifetime of the galaxy,
may fail to reproduce the chemical properties of Tucana. 
However, Hidalgo et al. (2013) computed simple CEMs, trying to reproduce the general behaviour of the \textit{SFR}(t) vs \textit{Z}(t) for Tucana. 
They assumed closed-box, or infall, or outflow scenarios and they concluded "For the dSphs, simple outflow models provide adequate fits to the observations".

For the above reason,  we test simple scenarios, considering the \textit{SFR}(t) by H13, but now trying to reproduce the $M_{gas}^{Obs}$ at present time and $Z_*(5.5Gyr)$, the $Z_{Obs}$ at 5.5 Gyr. 
Based on these models, we will quantify the disagreement with the observations to find a combination of scenarios at various stages of the galaxy's evolution that reproduces each \textit{Z} value determined by Hidalgo et al (2013).

The tested scenarios are summarized in Table 2 and described below:

\begin {enumerate}
	\item CBM
	\begin {enumerate}
		\item To reproduce $M_{gas}^{Obs}$ the model needs a $M_{gas}(0)$ = $2.38 \times 10^{6}M_{\odot}$, which corresponds to a $f_{b}(0)=5.948\times  10^{-4}$, and predicts $Z(5.5$ Gyr$)=33.9 Z_{Obs}$.
		\item To reproduce $Z_{Obs}$ the model needs a $M_{gas}(0)$ = $1.73 \times 10^{7}M_{\odot}$, which corresponds to a $f_{b}(0)=43.3\times  10^{-4}$, and predicts $M_{gas}(13.5$ Gyr$) = 993.3 $ $M_{gas}^{Obs}$.
	\end {enumerate}
	
	None of the CBMs can reproduce both observational constraints. Furthermore, the initial baryonic fraction required by the models are absurd for cosmology, about 3 orders of magnitude lower than the cosmological baryonic fraction ($f_b^{cos}= 0.17$), $f_{b}(0) \sim 0.003 f_{b}^{cos}$ and $f_{b}(0) \sim 0.025 f_{b}^{cos}$ respectively.
	 
	\item PAM
	
	Assuming the cosmological baryonic fraction $f_{b}=0.17$, which corresponds to a $M_{gas}(0)$ = $6.8 \times 10^{8}M_{\odot}$, we compute the following models:
	\begin {enumerate}	
		\item To reproduce $M_{gas}^{Obs}$, the model requires a negative accretion efficiency, $a=-286.6$ and predicts $Z(5.5$Gyr$) = 28.8 Z_{Obs}$.
		\item To reproduce $Z_{Obs}$, the model requires a negative accretion efficiency, $a=-280.2$, and predicts $M_{gas}(13.5$ Gyr$) =  1043$ $M_{gas}^{Obs}$. 
	\end {enumerate}

None of the PAMs are physically plausible because a$<$0 implies an outflow of primordial material.

 	\item WMM
 	
 	Assuming the cosmological baryonic fraction $f_{b}=0.17$, which corresponds to a $M_{gas}(0)$ = $6.8 \times 10^{8}M_{\odot}$, we compute the following models:
	\begin {enumerate}
		\item To reproduce $M_{gas}^{Obs}$, the model requires a wind efficiency $w=286.1$, and predicts $Z(5.5$Gyr$) = 0.17 Z_{Obs}$.  
		\item To reproduce $Z_{Obs}$, none of the w values can be obtained by keeping positive values of the gas mass. 
	\end {enumerate}

	 None of the WMMs can reproduce both observational constraints.

	\item RYM
			With CBM we reproduce $M_{gas}^{Obs}$, but $Z(5.5$ Gyr$)= 34.2 Z_{Obs}$. In order to reproduce $Z$(5.5Gyr) this model needs a yield reduction, dividing the original yield by $e=34.2$ (see Eq. 19).
	\end {enumerate}
	
\subsubsection{What did we learn from the simple models?}

By implementing each of the simple models throughout the lifetime of the galaxy, we noticed the following general behaviors:

a) None of the simple models can reproduce both observational constrains, $M_{gas}^{Obs}$ and $Z_{Obs}$. 

b) The initial baryonic fraction required by the models is 3 orders of magnitude smaller than the cosmological baryonic fraction.

c) Using the cosmological baryonic fraction $f_{b}^{cos}$, PAM predicts a$<0$, which does not have physical sense.

d) In the CBMs, \textit{Z} increases considerably and $M_{gas}$ decreases, both as a function of the \textit{SFR}.

e) In the PAMs, \textit{Z} decreases and $M_{gas}$ increases, both considerably, due to the fact that accretion is proportional to the \textit{SFR}.

f) In the WWMs, \textit{Z} decreases slightly and $M_{gas}$ decreases considerably, both as a function of the \textit{SFR}, due to the well mixed wind is that proportional to the \textit{SFR}.

g) In the RYM, \textit{Z} decreases in the same proportion as the reduction of the yield.

\begin{table*} 
\caption{Input parameters and results of the simple models of Tucana.}
 \label{Table2}
 \begin{tabular}{@{}lcccccc}
  \hline
 Model 	& $f_b(0)$	   &  $a$ & $w$ & $e$ & $M_{gas}(13.5 Gyr)$ &	$Z(5.5 Gyr)$ \\
            &($10^{-4}$) & 	  &          &	      & ($10^4M_\odot$)       &	($10^{-3}$) \\
  \hline
 CBM		& 5.948 & 0 & 0 &	1.0 & 1.5 & 46.84 \\
 CBM 	      & 	$43.300$ 	& $0$		&$0$			&	$1.0$ 	&$1490.0$	 &	$1.38$\\
 PAM		& 	$1700.0$ 	& $-286.6$	&$0$			&	$1.0$	&$1.5$	 &	$39.72$\\
 PAM		& 	$1700.0$ 	& $-280.1$	&$0$			&	$1.0$	&$1564.5$	 &	$1.38$\\
 WWM 	& 	$1700.0$ 	& $0$		&$286.1$	&	$1.0$	&$1.5$	 & 	$0.23$\\
WWM 	& 	$1700.0$ 	& $0$		&$-$			&	$1.0$	&$-$	& 	$1.38$\\ 
 RYM		& 	$5.948$		& $0$		&$0$			&	$34.2$	&$1.5$ &	$1.38$\\  
  \hline
Obs.		& 	$1700^c$ & $-$ &$-$	& $-$ & $1.5^b$ &	$1.38\pm0.23^a$\\
 \hline
 \end{tabular}
  \medskip\\
Column 1: Model name. Columns 2, 3, 4 and 5: input parameters of models: initial baryonic fraction, accretion efficiency, wind efficiency and yield reduction factor, respectively. Columns 6 and 7: model output values: current gas mass and metallicity of the youngest stars.\\ 
(${a}$) $Z_{obs}$ by Hidalgo et al. (2013), see section 2.
(${b}$) $M_{gas}^{Obs}$ by Oosterloo et al. (1996). 
(${c}$) $f_{b}^{cos}$, cosmological baryonic fraction, by Ade et al. (2015).
\end{table*}

%--------------------------------------------------------------------------------------------------------------------------------
\subsection{Best CEMs for Tucana (Complex models)}

Based on the \textit{SFR(t)} and \textit{Z(t)} behaviors (see upper panel of Figs. 1 and 2 ) and section 4.1.1, we realize that the star formation and chemical histories of Tucana suggest different galactic processes in different epochs, as described below:
\begin {itemize}

	\item Between 0 and 0.5 Gyr, the average of the \textit{SFR} and \textit{Z} are constant. This earliest stage of the galactic formation is quite complex 			(e.g. Sawala et al. 2010) due to the important accretion and feedback events. As a first approximation, a CBM represents a good average of the 			
			processes that occurred during the first $\sim$ 0.5 Gyr.
	\item Between 0.5 and 2 Gyr, the \textit{SFR} is the most intense and \textit{Z} remains constant. This can be explained by a metal dilution by primordial 		
			gas or by metal rich winds.
	\item Between 2 and 4 Gyr, the \textit{SFR} decreases slightly and \textit{Z} shows a slight increase, as a CBM predicts.
	\item Between 4 and 6 Gyr, the \textit{SFR} is reduced to almost zero, while \textit{Z} increases dramatically, reaching its maximum value at $ t \sim 6 $Gyr. This can be explained only if $M_{gas}$ at $ t \sim 4$ Gyr is very low, as a consequence of a huge amount of gas ejected to the intergalactic medium. 	
	\item Between 6 and 13.5 Gyr,  the \textit{SFR} is null. Therefore, there are no new stars and values of $Z_{*}$.
\end {itemize}

From these five epochs, and the fact that none of the simple models were successful in reproducing both, the observed $M_{gas}$ and \textit{Z} between 0-13.5 Gyr, we build complex models that combine the simple scenarios in different epochs.

\subsubsection{ Metal dilution model}

Taking into account the different epochs observed in the galaxy (section 4.2), we describe the stages of metal dilution model, corresponding to the different galactic processes. This model requires: 
	\begin {enumerate}
		\item A closed box model between 0 Gyr and 0.5 Gyr, with $M_{gas}(0)$ = $1.02 \times 10^{7}M_{\odot}$, which corresponds to an initial baryonic fraction of $f_{b}=2.547\times 10^{-3}$.
		\item A metal dilution from 0.5 to 2 Gyr with $a=13$.
		\item A closed box model between 2 Gyr and 4.5 Gyr.
		\item A strong, well-mixed wind at 4.5 Gyr.
		\item A closed box model between 4.5 Gyr and 13.5 Gyr.
	\end {enumerate}

During the first 0.5 Gyr, we assume that the galaxy behaved as a closed box. 
To reproduce \textit{Z}(0.5Gyr), the CBM requires a $M_{gas}(0)$ = $1.02 \times 10^{7}M_{\odot}$, which corresponds to an $f_{b}=2.547\times 10^{-3} = 0.015 f_{b}^{cos}$. 
During this first half Gyr, PAM may be plausible, but the $M_{gas}(0)$ still needs to be lower. 

Between 0.5 and 2 Gyr, the primordial infall dilutes the metals, maintaining a constant \textit{Z}. 
Those metals are produced and ejected by the huge amount of stars in the period of maximum \textit{SFR}.

Between 2 and 4.5 Gyr, the \textit{SFR} continues but with less intensity. 
The  death rate of stars during this period increases the predicted \textit{Z} according to $Z_{*}(t)$. 
This means that in those 2.5 Gyrs, the galaxy has neither metal dilution nor metal loss. 

At t=4.5 Gyr, a strong galactic wind occurs and the galaxy loses 98.7\% of its gas, keeping $M_{gas}$= $2.8\times10^{5}M_{\odot}$. 
This huge reduction of $M_{gas}$ is required in order to explain the considerable increase of \textit{Z} with the lowest \textit{SFR} during 4.5 and 6 Gyr. 
Due to at $ t \sim 4.5$ Gyr the \textit{SFR} is low to produce a supernova-driving gas loss (Dekel \& Silk 1986, Mc Low \& Ferrara 1999), we believe that the gas loss might be caused by an interaction with a massive system. However, other processes might have occurred (see section 6).

Between 4.5 and 6 Gyr, the star formation continues using the remaining gas, and \textit{Z} reaches its maximum value at $ t \sim$ 6 Gyr.

Since the \textit{SFR}(5.5 Gyr) is low ($4.7\times10^{3}M_{\odot}Gyr^{-1}$), the accuracy of \textit{Z} at 5.5 Gyr may be poor, and consequently we should not consider this last point to constrain the CEM. 
If we only assume \textit{Z} for t$\leq$4.5, the CEM does not change at all before $ t \sim$ 5 Gyr, because the \textit{Z}(5.5 Gyr) only constrains the amount of gas lost through the galactic wind at 4.5 Gyr. 
Therefore, we cannot estimate the intensity of the galactic wind that stopped the \textit{SFR}.

 \begin{figure}
 	\begin{center}
 		\includegraphics[scale=0.47]{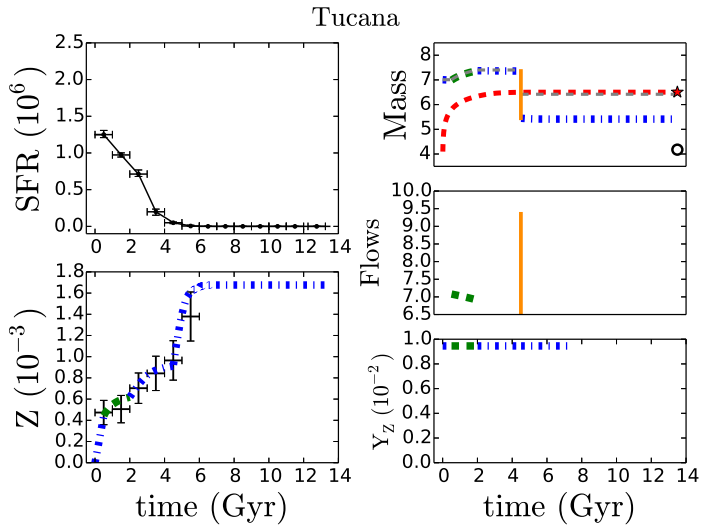}
 	\end{center}
 	\caption[Alphavsx]{\textbf{TUCANA. METAL DILUTION MODEL.}
 		\textbf{Top left:} Evolution of the star formation rate ($M_{\odot}$ Gyr$^{-1}$) considered in the model, \textit{SFR}(t). \textbf{Bottom left:} Evolution of gas metallicity, \textit{Z}(t). \textbf{Top right:} baryonic (dashed-thin grey line), formed star (dashed-thick red line), gaseous (blue, green and orange lines) mass evolution ($M_{\odot}$), in log. \textbf{Center right:} Evolution of gas flows ($M_{\odot} Gyr^{-1}$), in log. \textbf{Bottom right:} Required yield. \textbf{Epochs of the model:} Closed box (CBM, dotted-dashed blue line), primordial accretion (PAM, dashed-thicker green line), well mixed wind (WWM, continuous orange line). \textbf{Data:} \textit{SFR} and \textit{Z} by Hidalgo et al. 2013 (see section 2); star: mass of all stars ever formed, $M_{*}^{formed}$, (see section 2); and open circle: gaseous mass by Oosterloo et al. (1996).}
 	\label{fig03}
 \end{figure}
 %""""""""""""""""""""""""""""""""""""""""""""""""""""""""""""""""""""""""""""""""""""""""""""""""""""""" 	 

\subsubsection{ Metal loss model}

This model is identical to the metal dilution model except during the epoch (ii) (0.5-2 Gyr), when \textit{Z} is almost constant with a considerable \textit{SFR}. In this model, and during this period, the yield is reduced to reproduce the flat \textit{Z}.

Therefore this model requires: 

\begin {enumerate}
	\item A closed box model between 0 Gyr and 0.5 Gyr, with $M_{gas}(0)$ = $1.02 \times 10^{7}M_{\odot}$, which corresponds to an initial baryonic fraction of $f_{b}=2.547\times 10^{-3}$. 
	\item A reduced yield between 0.5 Gyr and 2 Gyr, with $Y_{Z}$ divided by $e=60.2$, see eq. 19.
	\item A closed box model between 2 Gyr and 4.5 Gyr.
	\item A strong, well-mixed wind at 4.5 Gyr.
	\item A Closed box model between 4.5 Gyr and 13.5 Gyr.
\end {enumerate}

In epoch (ii), the original yield was reduced by 98.3\% (e=60.2), meaning a loss of 98.3\% of the metals produced by the stars of the galaxy, this value is close to that calculated by Kirby et al. (2011b) (96\% to 99\%) for eight dSph galaxies of the Milky Way. 
These values are in agreement to that calculated by analytical and numerical models centered on the ejection efficiency of gas and metals in dwarf galaxies (Mc Low \& Ferrara 1999; Rodr\'iguez-Gonz\'alez et al. 2011, for example).

  \begin{figure}
 \begin{center}
   \includegraphics[scale=0.468]{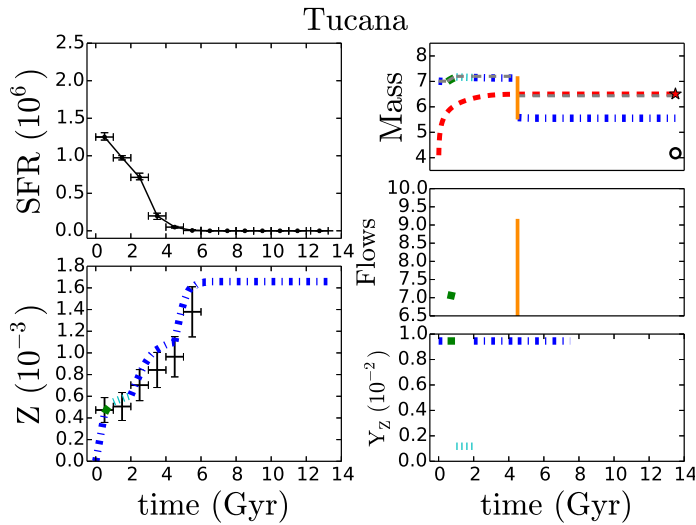}
 \end{center}
 \caption[Alphavsx]{\textbf{TUCANA. METAL LOSS MODEL.}
  
  \textbf{Panels and data:} As Figure 3. \textbf{Epochs of the model:} Closed box (CBM, dotted-dashed blue line), primordial accretion (PAM, dashed-thicker green line), well mixed wind (WWM, continuous orange line), reduced yield (RYM, dotted cyan line).}
 \label{fig04}
 \end{figure}
 
\subsubsection{Metal dilution \& metal loss models} 

The only difference between the metal dilution model and the metal loss model is the scenario that keeps \textit{Z} constant during a high \textit{SFR}, which occurs between 0.5 and 2.0 Gyr. 
To  discriminate between both models, we need good abundance determinations for chemical elements produced mainly by each type of star. 
That is, chemical elements synthesised by MS (e.g. Type II SNe) and other elements generated by single and binary LIMS (eg. AGBs and Type I SNe). Abundance ratios, of heavy elements produced by different stellar progenitors, do not change in a metal dilution model, but do change in a metal loss model. 
Therefore, we require models without IRA to know the delayed enrichment by LIMS and the amount of heavy metals possibly ejected into the intergalactic medium by SNe.

%"""""""""""""""""""""""""""""""""""""""""""""""""""""""""""""""""""""""""""""""""""""""""""""""""""""""

\subsection{Reproducing the current gas mass}	

The complex models, presented in the previous subsections, predict $M_{gas} \sim 1.8$ orders of magnitude higher than the observed value. 

To reduce the predicted $M_{gas}$ and match the $M_{gas}^{obs}$ we proposed two scenarios: 
i) a more efficient galactic wind at 4.5 Gyr (Fig. 5) without further loss of gas for  t$>$ 4.5 Gyr, 
or ii) a similar galactic wind of the previous models (sections 4.2.1 and 4.2.2) plus a gas loss for t$>$6  Gyr.

The first frame causes a loss of 98\% of the galactic gas, at t $\sim$ 4.5 Gyr, and predicts $Z(4.5$ Gyr)$\sim$ 6 $Z_{obs}$(4.5 Gyr). The dramatic rise of \textit{Z} is due to the fact that the same amount of stars enriches lower amounts of gas. In order to reproduce the metallicity of the youngest stellar generations (4.5$<$t(Gyr)$<$6), we infer a yield of 12\% (e=8.3) of its original value. Since the \textit{SFR} is very low during these 1.5 Gyrs, we conclude that the yield reduction is due to stochastic effects on the formation of massive stars (see Carigi \& Hernandez 2008). In Figure 5 we show this first scenario.

The second scenario considers that the difference in gas loss is due to external processes after the \textit{SFR} ends, to t$>$6  Gyr, like an interaction with a massive stellar system (Mayer et al. 2001). This scenario does not require either a stronger wind by SN heating or a yield reduction.

We consider that the two models present in section 4.2 are the best ones, because they explain most of the evolution of the galaxy (0-4.5 Gyr), although we cannot distinguish between the two scenarios of gas loss for $t>4.5$ Gyr described in the previous paragraphs. 
In order to discriminate between the two scenarios, for  t$>$ 4.5 Gyr, proposed in this section, a more precise and recent value of $M_{gas}^{Obs}$ is needed.

\begin{figure} 
  \begin{center}
    \includegraphics[scale=0.47]{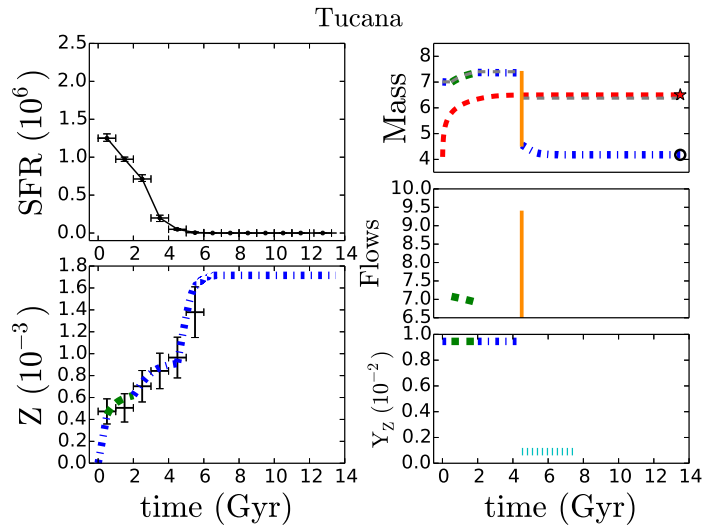}
  \end{center}
  \caption[Alphavsx]{\textbf{TUCANA. METAL DILUTION AND REDUCED YIELD MODEL TO REPRODUCE $M_{gas}^{Obs}$.} 
  
  \textbf{Panels and data:} As Figure 3. \textbf{Epochs of the model:} Closed box (CBM, dotted-dashed blue line), primordial accretion (PAM, dashed-thicker green line), well mixed wind (WWM, continuous orange line), reduced yield (RYM, dotted cyan line).}
  \label{fig05}
  \end{figure}
  
%------------------------------------------------------------------------------------------------------------------------------------------
\section{CEMs FOR CETUS}

Considering the similarities in the time-behavior of the \textit{SFR} and \textit{Z} in Cetus and Tucana (see Figs 1 and 2), we build similar combined models for Cetus as those developed for Tucana, assuming the same epochs, but with different accretion and galactic wind efficiencies. In this section we cannot calculate $f_b(0)$ because its virial mass has not been published in the literature.

%"""""""""""""""""""""""""""""""""""""""""""""""""""""""""""""""""""""""""""""""""""""""""""""""""""""""

\subsection{ Metal dilution model}

This model requires: 
	\begin {enumerate}
		\item A closed box model between 0 Gyr and 0.5 Gyr, with $M_{gas}(0)$ = $2.93 \times 10^{7}M_{\odot}$.
		\item A metal dilution from 0.5 to 2 Gyr with $a=20$.
		\item A closed box model between 2 Gyr and 4.5 Gyr.
		\item A strong, well-mixed wind at 4.5 Gyr.
		\item A closed box model between 4.5 Gyr and 13.5 Gyr.
	\end {enumerate}

During the first 0.5 Gyr, we assume that Cetus behaves as a closed box. 
Between 0.5 and 2 Gyr, the dwarf galaxy has a primordial accretion that increases the amount of gas, maintaining a constant \textit{Z}. 
Since Cetus presents higher $M_{gas}(0)$ and \textit{SFR} than Tucana, the model requires higher accretion. 
Between 2 and 4 Gyr, the \textit{SFR} continues with less intensity, until at $t=4.5$ Gyr a strong galactic wind occurs and the galaxy ejectes 98.6\% of its gas to the intergalactic medium, decreasing the gas mass to $M_{gas}$= $1.05\times10^{6}M_{\odot}$. 
Between 4.5 and  6 Gyr, with the remaining gas, the last stellar generation is formed and \textit{Z} reaches its maximum value at $ t \sim$ 6 Gyr.

\begin{figure}
\begin{center}
  \includegraphics[scale=0.46]{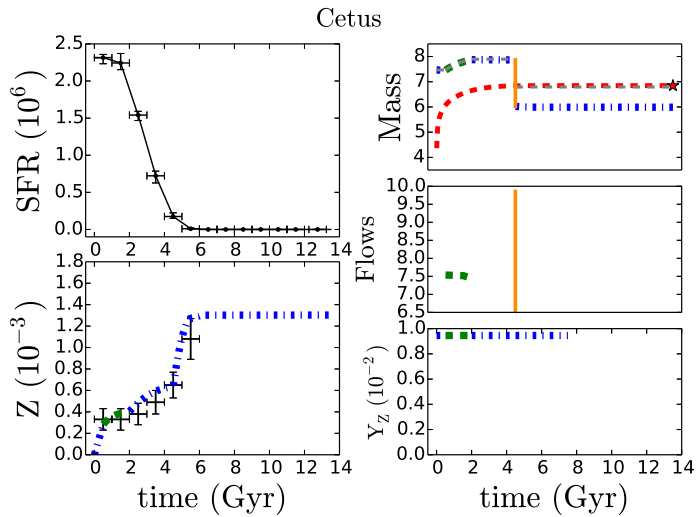}
\end{center}
	\caption[Alphavsx]{\textbf{CETUS. METAL DILUTION MODEL.}
 		
 		\textbf{Panels:} As Figure 3. \textbf{Epochs of the model:} Closed box (CBM, dotted-dashed blue line), primordial accretion (PAM, dashed-thicker green line), well mixed wind (WWM, continuous orange line). \textbf{Data:} \textit{SFR} and \textit{Z} by Hidalgo et al. 2013 (see section 2) and star: mass of all stars ever formed, $M_{*}^{formed}$ (see Section 2).}
\label{fig06}
\end{figure}

%"""""""""""""""""""""""""""""""""""""""""""""""""""""""""""""""""""""""""""""""""""""""""""""""""""""""

\subsection{ Metal loss model}

This model is identical to the metal dilution model except during the epoch (ii) (0.5-2 Gyr), when \textit{Z} is almost constant with a considerable \textit{SFR}. In this model, the yield is reduced to reproduce the flat \textit{Z}.

This model requires: 
\begin {enumerate}
\item A closed box model between 0 Gyr and 0.5 Gyr, with $M_{gas}(0)$ = $2.93 \times 10^{7}M_{\odot}$. 
\item A reduced yield between 0.5 Gyr and 2 Gyr, with $Y_{Z}$ divided by $e=149.8$.
\item A closed box model between 2 Gyr and 4.5 Gyr.
\item A strong, well-mixed wind at 4.5 Gyr.
\item A closed box model between 4.5 Gyr and 13.5 Gyr.
\end {enumerate}

In the epoch (ii) the original yield is reduced by 99.3\%, meaning a loss of 99.3\% of the metals produced by the stars in the galaxy. 
This percentage is in agreement with the value estimated by Kirby et al. (2011b) (96\% to 99\%) for other dSphs of the Milky Way.

\begin{figure}
\begin{center}
  \includegraphics[scale=0.46]{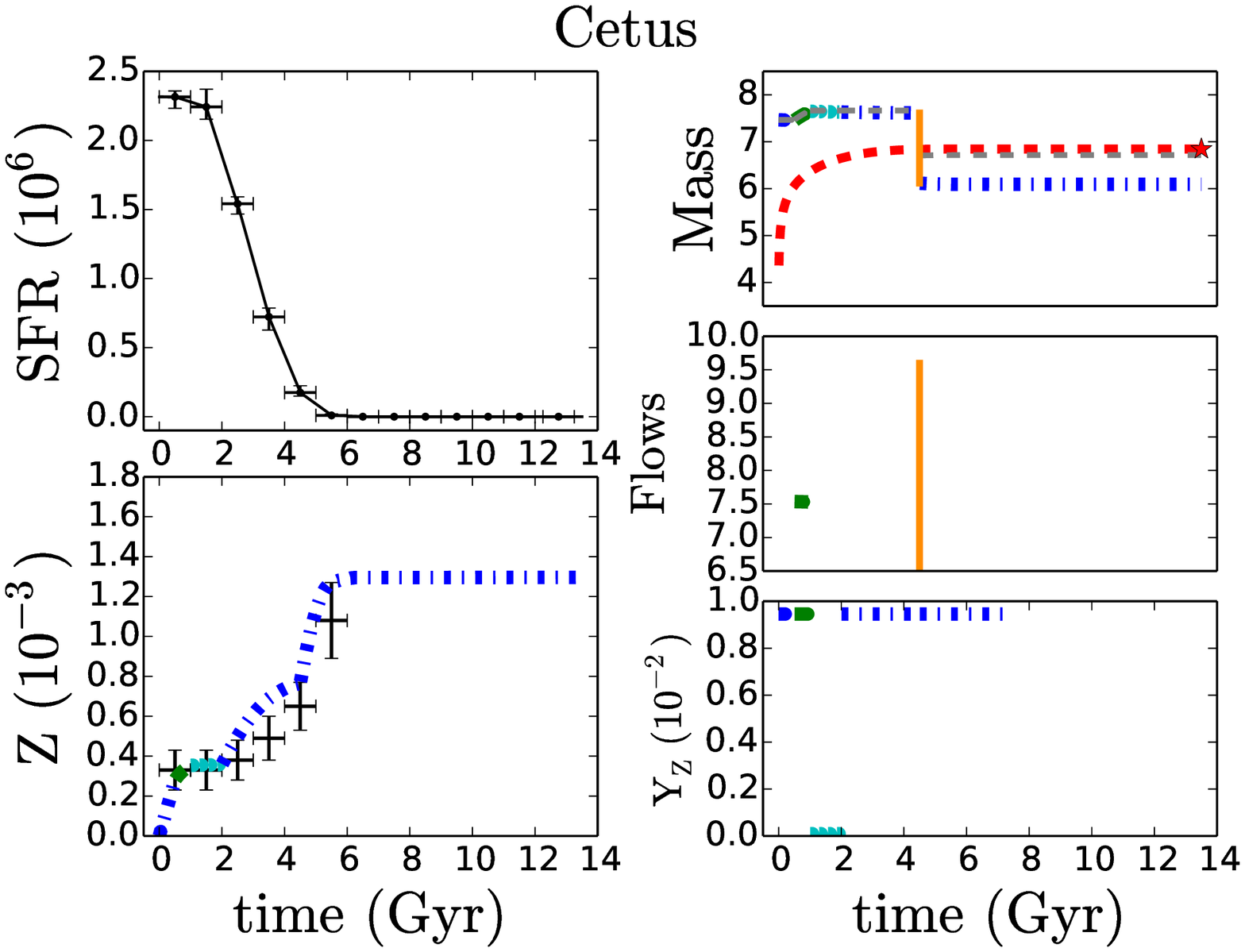}
\end{center}

 \caption[Alphavsx]{\textbf{CETUS. METAL LOSS MODEL.}
  
  \textbf{Panels and data:} As Figure 6. \textbf{Epochs of the model:} Closed box (CBM, dotted-dashed blue lines), primordial accretion (PAM, dashed-ticker green line), well mixed wind (WWM, continuous orange line), reduced yield (RYM, dotted cyan line).}

\label{fig07}
\end{figure}

\section{Environmental impact on the dSph evolution}

As can be seen in the Sections 4.2.1, 4.3 and 5.1:
i) the big reduction of \textit{SFR} and the high \textit{Z} values for $t \sim 4$ Gyr,
ii) the null \textit{SFR} for t $>$ 6 Gyr, and
iii) the gas lack at the present time, are explained by one or more episodes of gas loss.

Since the \textit{SFR} is very low or null in the last 9 Gyr, the gas loss of these dSphs would be caused by external agents, like 
gas stripping and tidal interactions due to a close encounter. 
Therefore, Tucana and Cetus probably did not evolve in absolute isolation, despite the fact that they have no close neighbouring galaxies at present. 
If those galaxies have been isolated during most of their evolution, then they have had few completed orbits inside the LG, so that their evolution is expected not to be strongly affected by giant galaxies (Monelli et al 2010a,b). 
Therefore  internal processes, like supernovae feedback, may originate the gas loss events required by the CEMs.
In order to understand the effects of the environment and internal processes in the evolution of Tucana and Cetus, 
reliable determinations of the orbits and the SFH of satellites and isolated dwarf galaxies  would be needed.
 
Moreover, in sections 4.2.2 and 5.2 the high \textit{SFR} and the ow \textit{Z} values, in the $0.5-2.0$Gyr range, are explained by a huge amount of metal expelled to the intergalactic medium and a negligible gas loss. 
These metals, together with the heavy elements ejected during the gas-loss episodes for t $>$ 4 Gyr, enriched the gas among the LG galaxies. Independently from metal-loss mechanism, the enrichment of the intergalactic medium due to the current dwarf galaxies may be partially unveiled when the star formation histories of these galaxies and chemical abundances of the intergalactic medium are determined.

\section{Discussion}

Based on the similarities between the time-behavior of the \textit{SFR} and \textit{Z} of Tucana and Cetus (see Figs. 1 and 2), 
two  isolated dSphs of the Local Group,
we focus the discussion on the comparison between our best models for each galaxy.
In Table 3 and Figures 8 and 9, we present together our best models, the metal dilution and metal loss models, respectively. 

One can notice that the time behavior of  $SFRs$ and ${Zs}$ are similar in both galaxies, 
but the $SFR$ of Cetus is higher, mainly during the first 5 Gyrs.
For example, at $t=0$ Gyr $SFR_{Cetus} \sim1.8 \ SFR_{Tucana}$. 
Consequently the formed stellar mass in Cetus is 2.2 higher (see Table 1) and the initial gas mass predicted by the Cetus models is 2.9 higher (see Table 3). 
Therefore, Cetus was less efficient in forming stars from the available gas, that is, $M_{star}/M_{gas}$(Cetus) = 0.8 $M_{star}/M_{gas}$(Tucana),
although Cetus is bigger than Tucana, that is, $r_h^{Cetus}= 2.5 r_h^{Tucana}= 703$pc,
being  $r_h$ the radius containing half the light of a galaxy (McConnachie 2012).

Since the $SFR_{Cetus} > SFR_{Tucana}$  and $Z_{Cetus} < Z_{Tucana}$, the accretion efficiency of Cetus is higher ($a_{Cetus}/a_{Tucana}=1.54$),  and the yield reduction of Cetus is higher ($e_{Cetus}/e_{Tucana}=2.5$).

Due to the $SFR(t)$, $M_{gas}(0)$ and $a$ values are higher in Cetus, the accretion rate, $A(t)$, and gas mass, $M_{gas}(t)$, are higher in the Cetus model (see Fig. 8). 
In particular, when the accretion ends, $A_{Cetus}$ (2 Gyr) = 3.4 $A_{Tucana}$ (2 Gyr), and at wind time, $ t_{wind}$,  $M_{gas}^{Cetus}$(4.5 Gyr) = 3.2 $M_{gas}^{Tucana}$(4.5 Gyr).

It is important to note that the outflow rate, \textit{W}(4.5Gyr), is higher in Cetus than in Tucana, because the metallicity increment rate, $\Delta Z/Z$, between 4.5 and 5.5 Gyr, is higher in Cetus. 
Specifically, $\Delta Z/Z_{Cetus}= 1.2 \Delta Z/Z_{Tucana}$. 
If these $Z$ values at 5.5 Gyr were not reliable, we could not compute $W$(4.5 Gyr) for both galaxies.

As we mentioned, the metal loss models are almost identical to the metal dilution models, except in the 0.5-2.0 Gyr period.
In this epoch, the metal loss models  consider yield reduction, instead of gas accretion, to explain the flat $Z$ when the $SFR$ is at its maximum.

Since the metal loss models do not include gas accretion, their $M_{gas}$ values are lower in the 2.0 - 4.5 Gyr range (see Fig. 9).
With a lower amount of gas and the same \textit{SFR}, the metallicities predicted by the metal loss models are higher (within the uncertainties) than those \textit{Zs} predicted by the metal dilution models. 
In addition, for the metal loss models the $M_{gas}$ is lower before the wind, and consequently the gas loss is lower at 4.5 Gyr. 
The evolution of the two types of models are identical after 4.5 Gyr.

 \begin{table} 
  \caption{Results of our best models for Tucana and Cetus.}
  \label{Table3}
  \begin{tabular}{@{}lcccccc}
   \hline
    & Tucana & Cetus\\
 \hline
 \hline
  Metal dilution model\\					
 $M_{gas}(0)$ $(10^{7}$ $M_{\odot})$		& 	$1.02$ 						& 	$2.93$	\\
 $t_{wind}$ (Gyr)									& 	$4.5\pm1.0$			            &  $4.5\pm1.0$\\
 \% $M_{lost}$ $_{gas}$									& 	$98.7$ 						& 	$98.6$	\\
 $Y_{Z}^{}/Y_{Z}^{original}$						& 	$1.00$ 						& 	$1.00$	\\
 \hline
  Metal loss model	\\
 $M_{gas}(0)$ $(10^{6}$ $M_{\odot})$		      & 	$1.02$ 						& 	$2.93$	\\
 $t_{wind}$ (Gyr)									& 	$4.5\pm1.0$		                  &  $4.5\pm1.0$\\
 \% $M_{lost}$ $_{gas}$							& 	$95.2$ 						& 	$94.3$	\\
 $Y_Z/Y_Z^{original}$	& 	$0.016$  &    	$0.006$\\
\hline
\hline
 \end{tabular}
  \medskip\\
Rows 1, 2, 3 and 4 for each model: Initial mass gas,
 time when the galactic wind occurs, 
 percentage of gas loss during the galactic wind and 
 required yield to original yield ratio.
\end{table}

\begin{figure} 
\begin{center}
\includegraphics[scale=0.47]{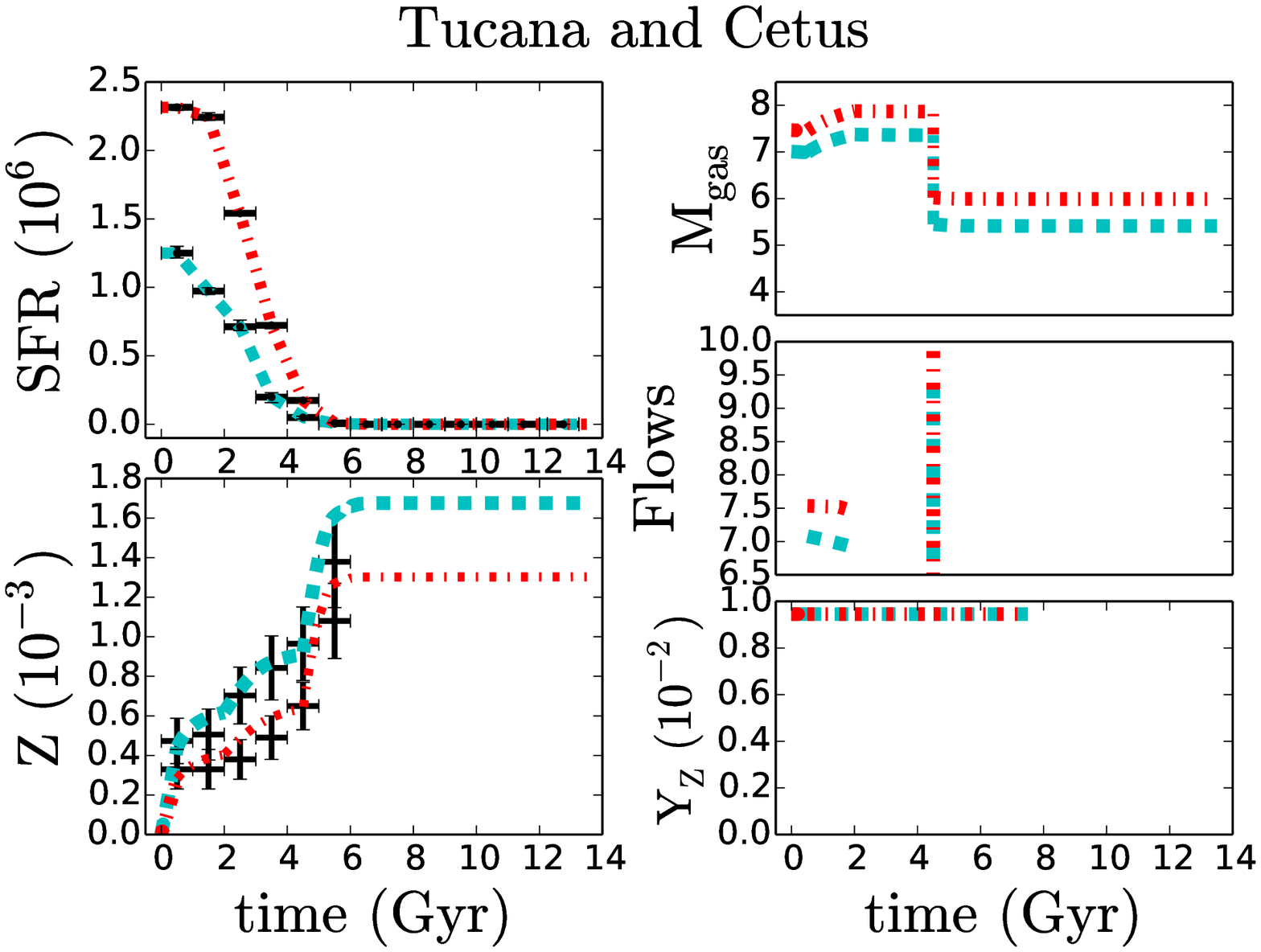}
\end{center}
\caption[Alphavsx]{\textbf{TUCANA AND CETUS. METAL DILUTION MODELS.}
    
\textbf{Top left:} Evolution of the star formation rate ($M_{\odot}Gyr^{-1}$) considered in the model, \textit{SFR}(time). \textbf{Bottom left:} Evolution of gas metallicity, \textit{Z}(time). \textbf{Top right:} Evolution of gas mass ($M_{\odot}$), in log. \textbf{Center right:} Evolution of gas flows ($M_{\odot}Gyr^{-1}$), in log. \textbf{Bottom right:} Required yield. \textbf{Lines:} Tucana (dashed cyan line) and Cetus (dotted-dashed red line). \textbf{Data:} \textit{SFR} and \textit{Z} by Hidalgo et al. 2013 (see section 2).}
\label{fig08}
\end{figure}

\begin{figure} 
\begin{center}
\includegraphics[scale=0.46]{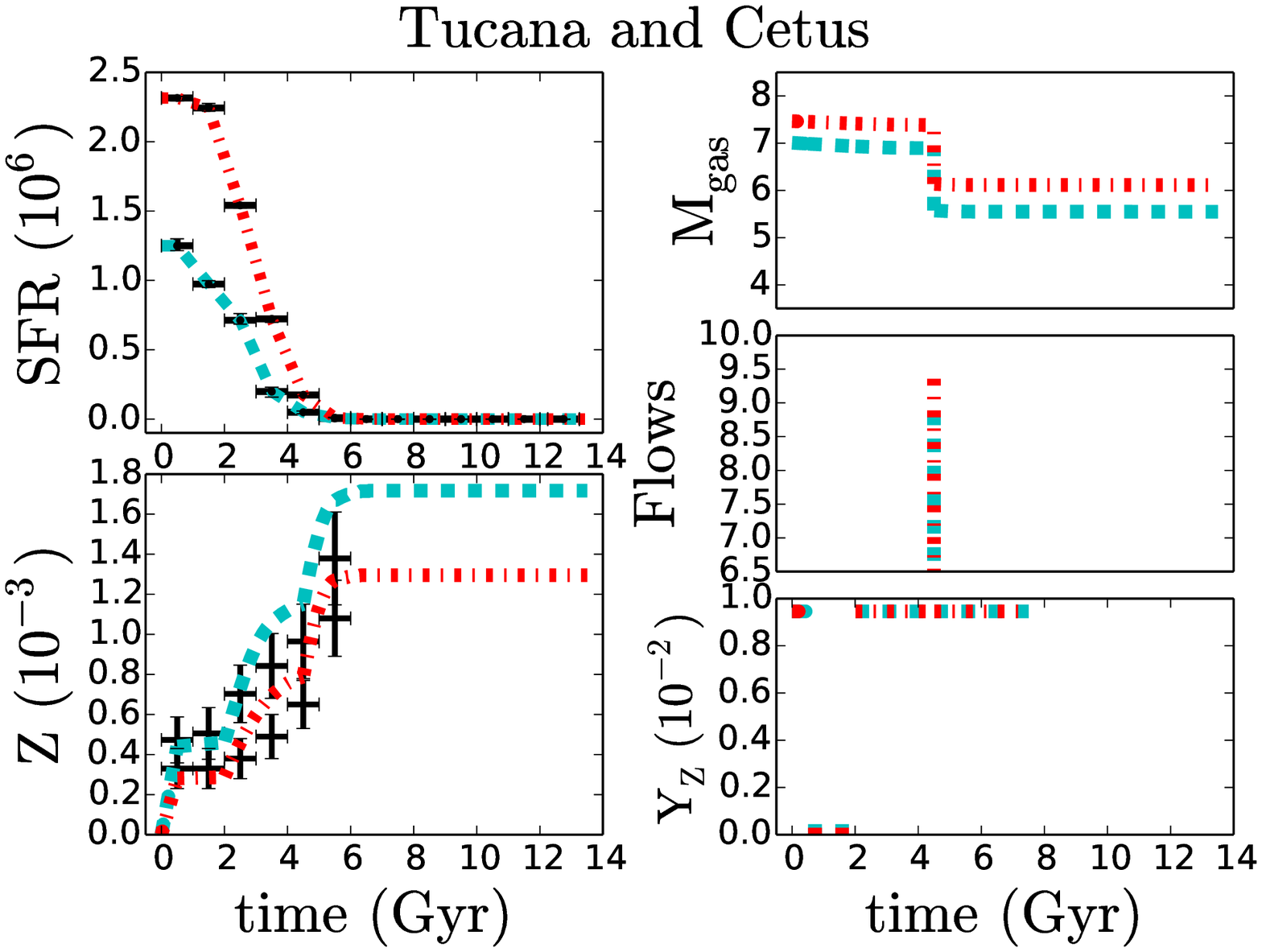}
\end{center}
\caption[Alphavsx]{\textbf{TUCANA AND CETUS. METAL LOSS MODELS.}

\textbf{Panels, lines and data:} As Figure 8.}
\label{fig09}
\end{figure}

%------------------------------------------------------------------------------------------------------------------------------------------
\section{Conclusions}

Based on the star formation and metallicity histories inferred from deep color-magnitude diagrams, we built accurate chemical evolution models for Tucana and Cetus.

We reached the following conclusions:

1) Even though Cetus is twice bigger and more massive than Tucana, the two galaxies had similar evolutions:

\begin {itemize}
\item During 75 \% of the star formation history, both galaxies behaved as closed boxes. 
\item During the remaining 25\%, specifically when the metallicity (\textit{Z}) was constant and the star formation rate (\textit{SFR})
was high, 
both galaxies received a high amount of primordial gas or lost metals.
\item During 0.1\% of the star formation history, previous to the SFR quenching, both galaxies lost most of their gas (94-99\%) to the intergalactic medium.
\end {itemize}

2) Each galaxy had a complex evolution, composed of different scenarios at different epochs:

\begin {itemize}
\item Between $0.5<t(Gyr)<2.0$, the constant $Z$ with the highest \textit{SFR} can be explained by two different scenarios: 
i) primordial accretion, which diluted the interstellar medium (ISM); or 
ii) metal-rich winds, which prevented the chemical enrichment of the ISM. 
To discriminate between these two scenarios, we need abundance ratios that are not available yet.

\item Between $2.0<t(Gyr)<4.5$, the \textit{Z} increased when the SFR decreased, according to a closed box model.

\item At $t \sim 4.5$ Gyr, a short-strong, well-mixed wind occurred and caused a huge reduction of the gas mass.

\item Between $4.5<t(Gyr)<6.0$, the \textit{SFR} was almost null and \textit{Z} increased abruptly, according to a closed box model with a low amount of gas mass. 

\item Between $4.5<t(Gyr)<6.0$, with the lowest \textit{SFR}, the massive-star formation would have been very low or null (stochastic effects in the IMF), which resulted in a yield decrease. 
Assuming a reduced yield, the amount of gas needed is even lower in order to explain the huge increase of \textit{Z}. 
\end {itemize}

3) Our models predict a current gas mass about only one order of magnitude lower than the current stellar mass. 
Observations of present-day gas mass are needed to constrain the amount of gas lost or gained after the star formation ceased. 

4) In both galaxies the star formation began with only 1.5\% of the baryonic mass fraction predicted by $\Lambda$CDM.

\section*{Acknowledgements}

We thank the anonymous Referee for the comments and suggestions that improved this paper. 
We acknowledge Vladimir Avila-Reese, Xavier Hernandez, Manuel Peimbert and Liliana Hern\'andez-Mart\'inez, for helpful suggestions. 
L.C. thanks for the financial supports provided by CONACyT of M\'exico (grant 241732), by PAPIIT of M\'exico  (IG110115, IA100815, IA101315) and by MINECO of Spain (AYA2010-16717). 
N. A. thanks for the financial support provided by Universidad  Iberoamericana.   

%$\big\}\Big\}\bigg\}\Bigg\}$
%\quad
%$\big\|\Big\|\bigg\|\Bigg\|$ 
%------------------------------------------------------------------------------------------------------------------------------------------
%\section{References}

%%%%%%%%%%%%%%%%
% BIBLIOGRAPHY %
%%%%%%%%%%%%%%%%
\bibliographystyle{mn2e}
\bibliography{Chemical-History-TucanaCetus}

\label{lastpage}

\end{document}